\documentclass{article}
\usepackage{amsmath,amsthm}
\newcommand{\pn}[1]{\left( #1 \right)}
\newtheorem{thm}{Theorem}
\DeclareMathOperator{\polylog}{polylog}
\usepackage[pdftex]{graphicx}

\begin{document}
\title{On the Complexity of the Monte Carlo Method for Incremental PageRank}
\author{Peter Lofgren
 \\
  Computer Science Department \\
  Stanford University\\
  plofgren@cs.stanford.edu}

\date{June 4, 2013}
\maketitle


\section{Abstract}
This note extends the analysis of incremental PageRank in [B. Bahmani, A. Chowdhury, and A. Goel. Fast Incremental and Personalized PageRank.  VLDB 2011].  In that work, the authors prove a running time of $O(\frac{nR}{\epsilon^2} \ln(m))$ to keep PageRank updated over $m$ edge arrivals in a graph with $n$ nodes when the algorithm stores $R$ random walks per node and the PageRank teleport probability is $\epsilon$.  To prove this running time, they assume that edges arrive in a random order, and leave it to future work to extend their running time guarantees to adversarial edge arrival.  In this note, we show that the random edge order assumption is necessary by exhibiting a graph and adversarial edge arrival order in which the running time is $\Omega \pn{R n m^{\lg{\frac{3}{2}(1-\epsilon)}}}$.  More generally, for any integer $d \geq 2$, we construct a graph and adversarial edge order in which the running time is $\Omega \pn{R n m^{\log_d(H_d (1-\epsilon))}}$, where $H_d$ is the $d$th harmonic number.


\section{Introduction}
In \cite{incremental_pagerank}, Bahmani, Chowdhury, and Goel propose a method of keeping an approximation to PageRank updated as edges from a graph arrive online.  They use the Monte Carlo method of computing PageRank \cite{monte_carlo_pagerank}.  In this method, we start at each node in the graph and take a random walk.  After each step of a random walk, we terminate the walk with probability $\epsilon$, the teleport probability, and it is complete.  With the remaining probability $1- \epsilon$, we transition to an out-neighbor of the current node, chosen uniformly at random, and continue the walk.  If we reach a node $v$ with outdegree 0 before completing the walk, we transition back to the starting node and continue from there.  To reduce variance, we take $R$ random walks per node, where $R$ might be a constant or $\log(n)$, depending on the accuracy required.  When a new edge $(u,v)$ is added to the graph, we consider revising each walk which passed through $u$, since it perhaps should have used this new edge.  The probability that the walk should have used this new edge is $\frac{1}{d(u)}$ where $d(u)$ is the new outdegree of $u$.  We flip a biased coin for each walk through $u$, and with probability  $\frac{1}{d(u)}$ we throw away the remainder of the walk and generate a new remainder starting with $v$.  To make sure that the length of each walk is geometrically distributed with expected length $\frac{1}{\epsilon}$, we preserve the length of the original walk when we generate a new remainder. 

When the graph is chosen by an adversary, but edges arrive in a random order, Bahmani, Chowdhury, and Goel \cite{incremental_pagerank} prove that the total work needed to keep an estimate of PageRank updated as $m$ edges arrive is $O \pn{\frac{nR}{\epsilon^2} \ln(m)}$, where $n$ is the number of vertices, $R$ is the number of stored walks per vertex, and $\epsilon$ is the teleport probability.   They state that it would be an interesting result to extend their running time guarantees to adversarial edge arrival.  This motivates the following question: does their algorithm require at most $O \pn{\frac{nR}{\epsilon^2} \ln(m)}$ total work for an adversarially chosen edge order? Our contribution is answering this question in the negative.

Note that we are not considering the absolute worst case performance of the algorithm in \cite{incremental_pagerank} or comparing it to other methods of computing PageRank.  We were motivated by our attempt to extend the bound of $\frac{nR}{\epsilon^2} \ln(m)$ work to the adversarial edge order model.  We discovered that such an extension is not possible.


\section{Result and Theory}
We first describe our construction in the case $d=2$ which corresponds to binary trees.
\begin{thm} Let $\epsilon < \frac{1}{3}$ be the teleport probability.  There exists a family of graphs with $n$ vertices and $m=n-1$ edges, and an edge order such that the total number of walk segments updated as the edges arrive is 
\[\Omega \pn{R n m^{\lg{\frac{3}{2}(1-\epsilon)}}}.\]
\end{thm}
For example if $\epsilon=.2$, the number of updates is $\Omega \pn{Rn m^{0.26}}$.  Hence the PageRank algorithm in \cite{incremental_pagerank} does not run in time  $O(R n \polylog(m))$ in the adversarial graph and adversarial edge order model.
\begin{proof} For any power of two, $N$, we describe how to construct a graph on $n = 2N-1$ nodes.  The case $N=16$ is shown in figure \ref{fig:graph},
  \begin{figure}[tbh]
    \centering
    \includegraphics[width=4in]{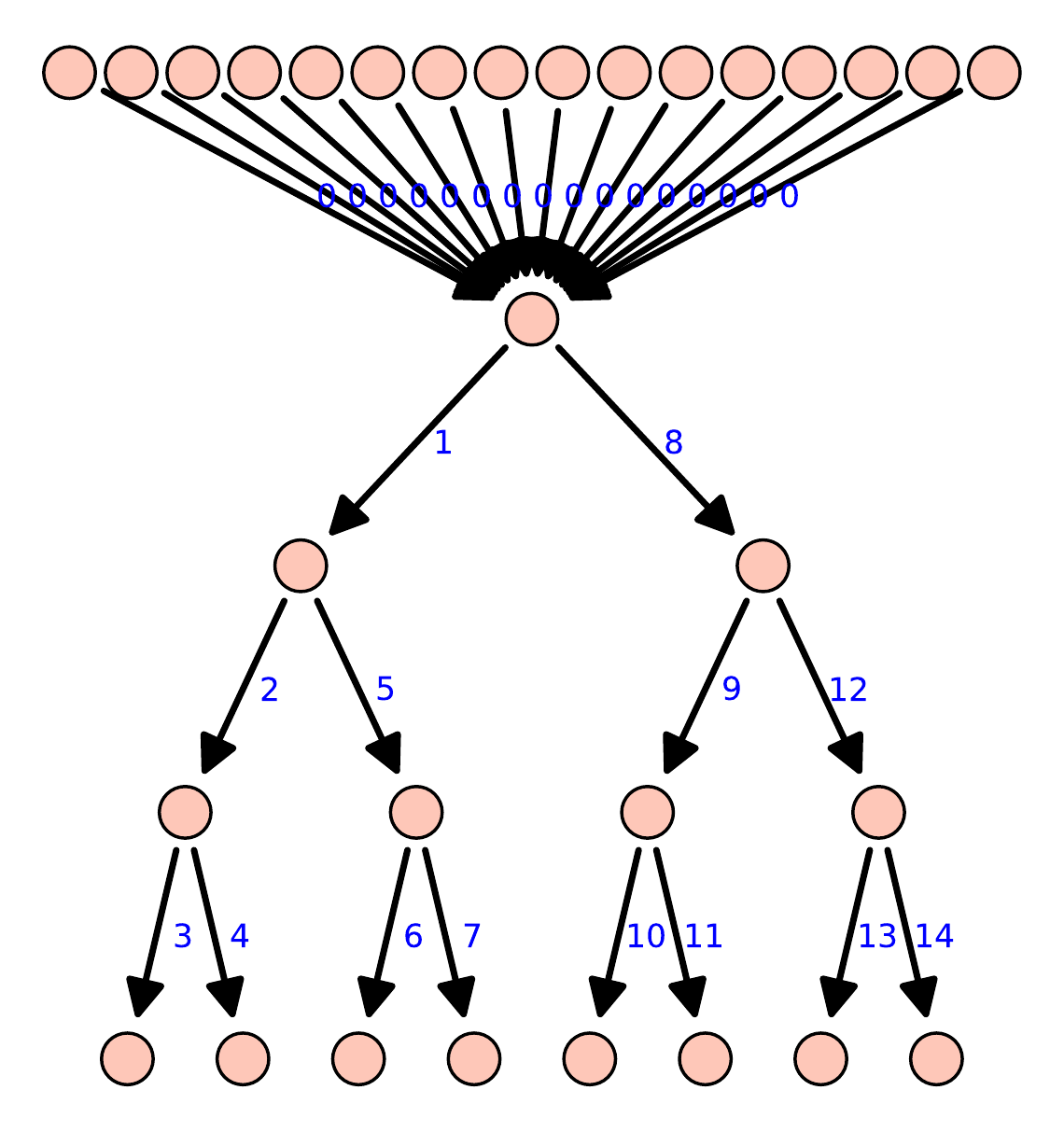}
    \caption{The case $N=16$ of our graph.  Edges are labeled with the order in which they arrive.}
    \label{fig:graph}
  \end{figure}
 with labels indicating the order of edge arrival.  The $n$ nodes are all present at the beginning, and the $m$ edges arrive one at a time.  There is a top row of $N$ nodes, each connected to the root of a balanced binary tree of $N-1$ nodes.  The edges of the top row arrive first, creating $RN$ walk segments to the root of the binary tree.  The edges in the tree arrive in a depth-first traversal of the binary tree starting at the root, so PageRank is funneled toward a leaf before being diluted among the branches.  The left edge leaving each vertex arrives before the right edge, so when the left edge leaving a vertex $u$ arrives, any incomplete walk through $u$  will need to be updated.  When the right edge leaving a vertex $u$ arrives, any incomplete walk will need to be updated with probability $\frac{1}{2}$.  Consider the probability that a walk from the root needs to be updated when an edge $(u,v)$ arrives to a node $u$ in row $i$.  First of all the walk needs to have length at least $i$ which happens with probability $(1-\epsilon)^i$.  Now as we trace the unique path from the root to $u$, the probability that a walk follows this path is the probability that it takes the correct edge (right or left) leaving each vertex.  Because the left edge to each vertex on any path from the root arrives before the right edge, the walk is guaranteed to follow the path to $u$ at left edges, while at right edges it has probability $\frac{1}{2}$ of going towards $u$.  Thus if there are $k$ right edges on the path and $i-k$ left edges, the probability that a walk of length at least $i$ will reach $u$ is $\pn{\frac{1}{2}}^k$.   Since there are $RN$ paths from top nodes which could potentially reach $u$, the expected number of paths which need to be updaded when edge $(u,v)$ arrives is
\[R N(1-\epsilon)^i  \left(\frac{1}{2}\right)^k .\]

In the $i$th row, there are $\binom{i}{k}$ nodes which can be reached via $k$ right branches and $i-k$ left branches from the root.  Thus the total expected number of path segments updated due to edges in the $i$th row is
\[R N(1-\epsilon)^i \sum_{k=0}^i \binom{i}{k} \left(\frac{1}{2}\right)^k = R N \left(\frac{3}{2}(1-\epsilon)\right)^i \]
after applying the binomial theorem.
 There are $\lg(N)$ rows in the tree, so the total number of path segments updated is
\begin{align*}
 R N \sum_{i=0}^{\lg(N)-1}  \pn{\frac{3}{2}(1-\epsilon)}^i &= R N \frac{\pn{\frac{3}{2}(1-\epsilon)}^{\lg{N}}-1}{\frac{3}{2} (1-\epsilon)-1}  \\
&= R N \frac{N^{\lg{\pn{\frac{3}{2}(1-\epsilon)}}}-1}{\frac{3}{2} (1-\epsilon)-1}  \\
\end{align*}
Now using the relations $n=2N-1 = \Theta(N)$ and $m=2N-2 = \Theta(N)$, the result follows.
\end{proof}

The above theorem needed to assume $\epsilon < \frac{1}{3}$, because for larger teleport probabilities, the walks from the top row of nodes will terminate on average before they can reach a significant number of the nodes in the tree, so the walks in this construction can be updated efficiently in time $O(R n \polylog(m))$ time.  To disprove the conjecture in the case $\epsilon \geq \frac{1}{3}$, we  generalize the above result to $d$-ary trees so walks from the top row can reach a larger number of tree nodes as the edges arrive.

\begin{thm} Let $\epsilon$ be the teleport probability.  For each branching factor $d$ such that $H_d (1-\epsilon) > 1$ , there exists a family of graphs where $m=n-1$ and an edge order such that the total number of walk segments updated as the edges arrive is 
\[\Omega \pn{Rn m^{\log_d(H_d (1-\epsilon))}}\]
where $H_d$ is the $d$th harmonic number.
\end{thm}
For any $\epsilon$, if we set $d$ such that $H_d (1-\epsilon) > 1$, we see that this running time is greater than
$O(Rn \polylog(m))$.  

\begin{proof}
  Given $N$, we describe how to construct a graph on $n=2N$ nodes.  We construct a graph similar to the binary construction above, but place $d$ children under each tree node.  Let the $N$ top edges arrive first, and the remaining edges arrive in a depth-first traversal of the tree.  Consider the probability that a random walk needs to be updated when an edge $(u, v)$ arrives in row $i$.  On the unique path from the root to $u$, let $j_r \in \{1,\ldots, d\}$ be the index of the child of the node followed in row $r$.  The probability that a random path will follow the branch toward $u$ in row $r$ is $\frac{1}{j_r}$, since at the time $u$ arrives, $u$'s ancestor in row $r$ will have exactly $j_r$ children.  In addition, to reach node $u$, a random path from the root must have  length at least $i$, which happens with probability $(1-\epsilon)^i$.  Thus the probability that a random path from the root is updated when edge $(u,v)$ arrives is \[(1-\epsilon)^i \prod_{r=1}^i \frac{1}{j_r}.\]
Note that this generalizes the binary case, where we let $k$ be the number of rows $r$ such that $j_r=2$.  Summing over all nodes $u$ in row $i$ is equivalent to varying all indices $j_r$ within their ranges.  Thus the expected number of paths updated as edges in row $i$ arrive is
\[RN \sum_{j_1=1}^d \sum_{j_2=1}^d \cdots \sum_{j_i=1}^d (1-\epsilon)^i \prod_{r=1}^i \frac{1}{j_r} = RN \pn{(1- \epsilon)H_d }^i \]
where $H_d$ is the $d$th harmonic number.
Since there are $N$ nodes in the tree, there are at least $\lfloor \log_d(N) \rfloor$ rows.  Thus the total expected number of updates as the $N-1$ tree edges arrive is at least
\begin{align*}
 R N \sum_{i=0}^{\log_d(N)-1}  \pn{H_d(1-\epsilon)}^i &= R N \frac{\pn{H_d(1-\epsilon)}^{\log(N)/\log(d)}-1}{H_d (1-\epsilon)-1}  \\&= \Theta \pn{RN^{1+\log(H_d (1-\epsilon))/\log(d)}}.
\end{align*}
Now using the relations $n=2N = \Theta(N)$ and $m=2N-1 = \Theta(N)$, the result follows.
\end{proof}
\section{Acknowledgments}
Thanks to Prof.~Ashish Goel for suggesting this problem and to Rishi Gupta for helpful conversation.  I was supported by the National Defense Science and Engineering Graduate Fellowship (NDSEG) Program.

\end{document}